\documentstyle[12pt,lathuile]{article}

\newcommand{\lbl}[1]{\label{eq:#1}}
\newcommand{ \rf}[1]{(\ref{eq:#1})}

\newcommand{\be}{\begin{equation}}
\newcommand{\ee}{\end{equation}}
\newcommand{\bea}{\begin{eqnarray}}
\newcommand{\eea}{\end{eqnarray}}
\newcommand{\setl}{\setlength\arraycolsep{2pt}} 

\newcommand{\noi}{\noindent}
\newcommand{\nn}{\nonumber}
\newcommand{\ra}{\rightarrow}

\newcommand{\lesssim}{ {\
\lower-1.2pt\vbox{\hbox{\rlap{$<$}\lower5pt\vbox{\hbox{$\sim$}}}}\ } 
}
\newcommand{\gtrsim}{ {\
\lower-1.2pt\vbox{\hbox{\rlap{$>$}\lower5pt\vbox{\hbox{$\sim$}}}}\ } 
}

\newcommand{\cF}{{\cal F}}

\newcommand{\cH}{{\cal H}}

\newcommand{\cO}{{\cal O}}

\newcommand{\cW}{{\cal W}}

\newcommand{\Imm}{\mbox{\rm Im}}

\newcommand{\MeV}{\mbox{\rm MeV}}
\newcommand{\GeV}{\mbox{\rm GeV}}

\newcommand{\annd}{\mbox{\rm and}}

\newcommand{\GF}{G_{\mbox{\rm {\tiny F}}}}

\input epsf

\begin{document}

\title{ 
THE MUON $g-2$ REVISITED\\\vspace*{0.25cm}
{\it Invited talk at the XVI Rencontres de Physique de La Vallée d'Aoste
}}
\author{
Eduardo de Rafael       \\
{\em CPT, CNRS--Luminy, Marseille} \\}
\maketitle
\baselineskip=14.5pt
\begin{abstract}
I present a short review of the present status of the  
Standard Model prediction of the anomalous magnetic moment of the muon,
with special emphasis on the hadronic contributions. 
\end{abstract}
\baselineskip=17pt
\newpage

%%%%%%%%%%%%%%%%%%%%%%
\section{Introduction}
%%%%%%%%%%%%%%%%%%%%%%

\noi
The $g$--factor of the muon is the quantity which relates its spin
$\vec{s}$ to its magnetic moment $\vec{\mu}$ in appropriate units:
\be
\vec{\mu}=g_{\mu}\frac{e\hbar}{2m_{\mu}c}\vec{s}\,,\qquad\annd\qquad
\underbrace{g_{\mu}=2}_{\mbox{\rm \small
Dirac}}(1+a_{\mu})\,.
\ee
In the Dirac theory of a charged spin--$1/2$ particle, $g=2$. Quantum
Electrodynamics (QED) predicts deviations from the Dirac prediction,
because in the presence of an external magnetic field the muon
(electron) can emit and reabsorb virtual photons. The correction
$a_{\mu}$ to the Dirac prediction is called the anomalous magnetic moment.
It is a quantity directly accessible to experiment \footnote{See e.g.
ref.\cite{Far01} for a simple and lucid review where references to the
experimental literature can also be found.}.

The experimental world average, at the time of the La Thuile meeting,
which included the BNL published result\cite{BNL} based on the 1999
$\mu^{+}$ data, was
\be\label{BNL}
a_{\mu}(\mbox{\rm exp.})=11~659~202.3(15.1)\times
10^{-10}\  {[1.3\ {\mbox{\rm ppm}}]}\,.
\ee
There is a recent new result from the BNL collaboration\cite{BNL1},
based on
$\mu^{+}$ data collected in the year 2000, 
\be\label{BNL1}
a_{\mu}(\mbox{\rm exp.})=11~659~204(7)(5)\times
10^{-10}\  {[0.7\ {\mbox{\rm ppm}}]}\,.
\ee
With this result, the present world average is now
\be\label{wa}
a_{\mu}(\mbox{\rm exp.})=11~659~203(8)\times
10^{-10}\  {[0.7\ {\mbox{\rm ppm}}]}\,.
\ee

In this talk, I
shall present a review of the various contributions to
$a_{\mu}$ in the Standard Model, with special emphasis in a recent
evaluation of the dominant contribution from the hadronic 
light--by--light scattering\cite{KNb01}, which has the merit to have
stopped an avalanche of theoretical speculations, at least temporarily.

%%%%%%%%%%%%%%%%%%%%%%%%%%%%%%%%%%%%%%%%%%%%%%%
\section{Some Remarks on the QED Contributions}
%%%%%%%%%%%%%%%%%%%%%%%%%%%%%%%%%%%%%%%%%%%%%%%

\noi
In QED, the Feynman diagrams which contribute to $a_{\mu}$ at a given
order in the perturbation theory expansion (powers of
$\frac{\alpha}{\pi}$), can be classified in four classes:
\goodbreak
\begin{description}
\item[]{\it Diagrams with virtual photons and muon loops }

Examples of that are the lowest order contribution in Fig.~1 and the
two loop contributions in Fig.~2. In full generality, this is the class of
diagrams which includes those with only virtual photons, and the ones
where the internal fermionic lines are of the same flavour as the
external line. Since
$a_{\mu}$ is a dimensionless quantity, these contributions are purely
numerical, and they are the same for the three charged leptons:
$l=e\,,\mu\,,\tau$. Indeed,
$a_{l}^{(2)}$ from Fig.~1 is the celebrated Schwinger result\cite{Sch48}
\be\label{sch}
a_{l}^{(2)}=\frac{1}{2}\frac{\alpha}{\pi}\,,
\ee

%%%%%%%%%%%%%%%%%%%%%%%%%%%%%%%%%%%%%%
\vskip 1pc
\centerline{\epsfbox{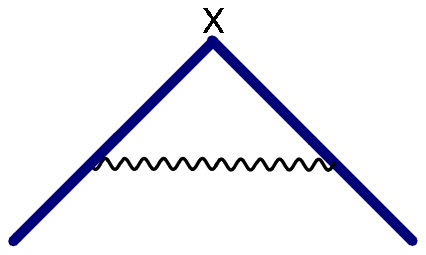}}
\begin{center}
{{\bf Fig.~1} {\it Lowest Order QED Contribution}}
\end{center}
%\vskip 1pc
%%%%%%%%%%%%%%%%%%%%%%%%%%%%%%%%%%%%%%

%%%%%%%%%%%%%%%%%%%%%%%%%%%%%%%%%%%%%%
\vskip 1pc
\centerline{\epsfbox{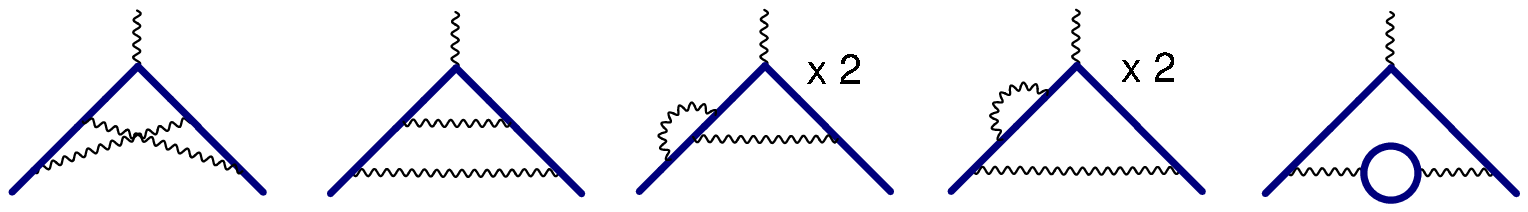}}
\begin{center}
{{\bf Fig.~2} {\it  QED Contribution at the Two Loop Level}}
\end{center}
%\vskip 1pc
%%%%%%%%%%%%%%%%%%%%%%%%%%%%%%%%%%%%%%

while $a_{l}^{(4)}$ from the seven diagrams in Fig.~2 gives the
result\cite{Pet57,Som57} 
\be
a_{l}^{(4)}
={\left\{ 
\frac{197}{144}+\frac{\pi^2}{12}-\frac{\pi^2}{2}\ln
2+\frac{3}{4}\zeta(3)\right\}}\left(\frac{\alpha}{\pi}\right)^2\,.
\ee

At three loops there are $72$--Feynman diagrams of this type which
contribute. Quite remarkably, they are also known
analytically\cite{LR96}. They bring in transcendental numbers like
$\zeta(3)$, the Riemann zeta--function of argument $3$, and of higher
complexity.

At the four loop
level, there are 891 Feynman diagrams of this type,
and their numerical evaluation is still in progress\cite{Kin03}. 

\item[]{\it Vacuum Polarization Diagrams from Electron Loops}

The simplest example is the Feynman diagram in Fig.~3, which gives a
contribution
\be\label{vpe}
a_{\mu}=\Big[\underbrace{\left({\frac{2}{3}}
\right)}_{\beta_{1}}
\left({\frac{1}{2}}\right)
\log\frac{m_{\mu}}{m_{e}}-\frac{25}{36}
+\cO\left(
\frac{m_{e}}{m_{\mu}}\right)
\Big]\left(\frac{\alpha}{\pi} \right)^2\,.
\ee
These contributions are enhanced by QED short--distance logarithms of the
ratio of the muon mass to the electron mass, and are therefore very
important. As shown in ref.\cite{LdeR74}, they are governed by a
Callan--Symanzik type equation
\be\label{cs}
\left(m_{e}\frac{\partial}{\partial
m_{e}}\!+\!\beta(\alpha)\alpha\frac{\partial}{\partial\alpha}
\right)\!a_{\mu}^{(\infty)}(\frac{m_{\mu}}{m_{e}},\alpha)\!=\!0\,,
\ee
where $\beta(\alpha)$ is the QED--function associated with charge
renormalization, and $a_{\mu}^{(\infty)}(\frac{m_{\mu}}{m_{e}},\alpha)$
denotes the contribution to $a_{\mu}$ from powers of logarithms and
constant terms. This renormalization group equation is at the origin of
the simplicity of the result in Eq.(\ref{vpe}). The factor $2/3$ in front
of $\log\frac{m_{\mu}}{m_{e}}$ comes from the first term in the
$\beta$--function and the factor $1/2$ is the lowest order coefficient of
$\alpha/\pi$ in Eq.(\ref{sch}), which fixes the boundary
condition to solve the differential equation in (\ref{cs}) at the first
non--trivial order in perturbation theory i.e., 
$\cO(\frac{\alpha}{\pi})^2$. 
%%%%%%%%%%%%%%%%%%%%%%%%%%%%%%%%%%%%%%
\vskip 1pc
\centerline{\epsfbox{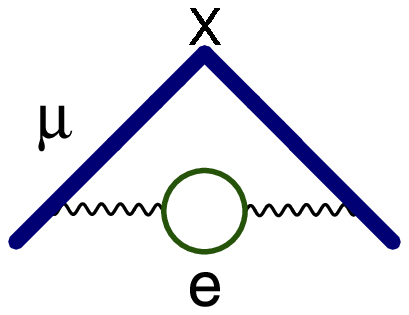}}
\begin{center}
{{\bf Fig.~3} {\it  Vacuum Polarization contribution from a Small Internal
Mass}}
\end{center}
%\vskip 1pc
%%%%%%%%%%%%%%%%%%%%%%%%%%%%%%%%%%%%%%
Knowing the QED $\beta$--function at
three loops and $a_{\mu}$ (from
the universal class of diagrams discussed above) also at three loops,
allows one to sum leading, next--to--leading, and
next--to--next--to--leading powers of
$\log m_{\mu}/m_e$ to all orders in perturbation theory. Of course, these
logarithms can be reabsorbed in a running fine structure coupling
$\alpha(m_{\mu})$. It is often forgotten that the first
experimental evidence for the running of a coupling constant in quantum
field theory comes in fact from the anomalous magnetic moment of the muon
in QED, well before QCD and well before the measurement of
$\alpha(M_{Z})$. 

\item[]{\it Vacuum Polarization Diagrams from Tau Loops} 

The simplest example is the Feynman diagram in Fig.~4 below, 
%%%%%%%%%%%%%%%%%%%%%%%%%%%%%%%%%%%%%%
\vskip 1pc
\centerline{\epsfbox{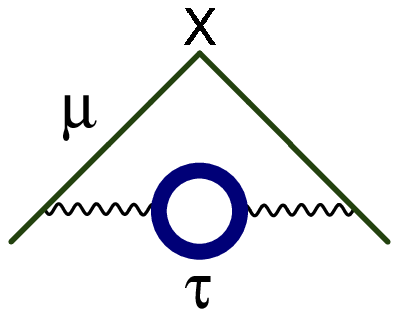}}
\begin{center}
{{\bf Fig.~4} {\it  Vacuum Polarization contribution from a Large Internal
Mass}}
\end{center}
%\vskip 1pc
%%%%%%%%%%%%%%%%%%%%%%%%%%%%%%%%%%%%%%

which gives a
contribution
\be
\alpha(m_{\mu})=
{\left[\frac{1}{45}\left(\frac{m_{\mu}}{m_{\tau}}\right)^2+\cO\left(
\frac{m_{\mu}^4}{m_{\tau}^4}\log\frac{m_{\tau}}{m_{\mu}}\right)
\right]}\left(\frac{\alpha}{\pi} \right)^2\,.
\ee
In full generality, internal heavy masses in the vacuum polarization
loops (heavy with respect to the external leptonic line) decouple. 

\item[]{\it Light--by--Light Scattering Diagrams from Electron Loops}

It is well known that the light--by--light diagrams in QED are
convergent, (once the full set of gauge invariant combinations is
considered). Because of that, it came as a big surprise to find out that
the set of diagrams in Fig.~5,
%%%%%%%%%%%%%%%%%%%%%%%%%%%%%%%%%%%%%%
\vskip 1pc
\centerline{\epsfbox{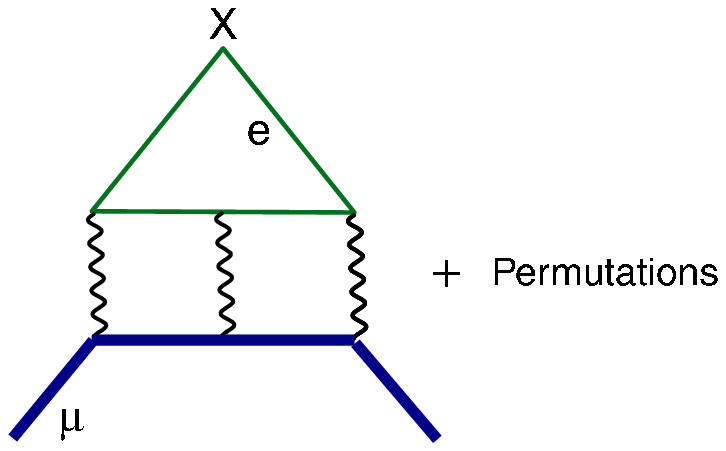}}
\begin{center}
{{\bf Fig.~5} {\it Light--by--Light Scattering contribution from a Small
Internal Mass}}
\end{center}
%\vskip 1pc
%%%%%%%%%%%%%%%%%%%%%%%%%%%%%%%%%%%%%%
when the lepton in the loop has a mass
smaller than the external leptonic line, produces a contribution
proportional to
$\log(m_{\mu}/m_{e})$; and, in fact, with a large
coefficient\cite{Kinetal69}
\be
a_{\mu}^{(3)}\vert_{\mbox{\tiny\rm
l.byl.}}=\left[\frac{2}{3}\pi^2\ln\frac{m_{\mu}}{m_{e}}
+\cdots\right]\left(\frac{\alpha}{\pi}\right)^3=20.947...
\left(\frac{\alpha}{\pi}\right)^3\,.
\ee

This contribution is now known, analytically, for arbitrary
values of the lepton masses\cite{LR93}. It can be understood in the
framework of effective field theories with different scales of masses.
Again, it can be easily shown that internal leptonic loops with a heavy
mass compared to the external leptonic line, decouple.  

\end{description}

\noi
Altogether, the purely QED contribution to the muon anomalous magnetic
moment of the muon, including $e\,,\mu\,,$ and $\tau$ lepton loops 
is known to an accuracy which is certainly good enough for the present
comparison between theory and experiment 
\be\label{qed}
a_{\mu}(\mbox{\rm\tiny {QED}})=(11~658~470.57\pm 0.29)\times
10^{-10}\,.
\ee
This is the number one gets, using the determination of the
fine structure constant
\be
\alpha^{-1}=137.035~999~59(52) [3.8~\mbox{\rm ppb}]\,,
\ee
which follows from the comparison between the
experimental determination of the electron (positron) anomalous magnetic
moments\cite{VD87} and the QED theoretical prediction (see e.g.
ref.\cite{MT00} and references therein). The error in Eq.(\ref{qed}) is
mostly due to the experimental errors in the determination of the ratios
of lepton masses and the error in the numerical integration of some of
the four--loop contributions. 

The
question which naturally arises is whether or not the discrepancy between
the experimental numbers in Eqs.(\ref{BNL}) and (\ref{BNL1}) on the one
hand and the QED contribution from leptons alone, can be understood in
terms of the extra hadronic and electroweak contributions predicted by the
Standard Model. 

%%%%%%%%%%%%%%%%%%%%%%%%%%%%%%%%%%%%%%
\section{Hadronic Vacuum Polarization}
%%%%%%%%%%%%%%%%%%%%%%%%%%%%%%%%%%%%%%

\noi
This is the contribution illustrated by the diagram in Fig.~6, with 
the shade in the vacuum polarization indicating hadrons. 
%%%%%%%%%%%%%%%%%%%%%%%%%%%%%%%%%%%%%%
\vskip 1pc
\centerline{\epsfbox{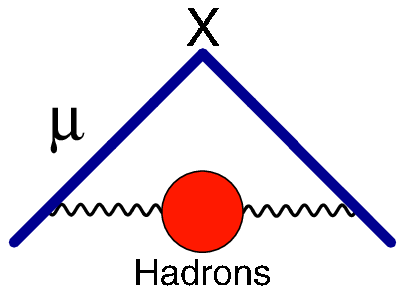}}
\begin{center}
{{\bf Fig.~6} {\it Hadronic Vacuum Polarization Contribution}}
\end{center}
%\vskip 1pc
%%%%%%%%%%%%%%%%%%%%%%%%%%%%%%%%%%%%%%
\noi
All the estimates
of this contribution are based on the spectral representation
\be\lbl{specrep}
a_{\mu}^{\mbox{\rm\tiny
(h.~v.p.)}}=\frac{\alpha}{\pi}\int_{0}^{\infty}\frac{dt}{t}
{\frac{1}{\pi}\Imm\Pi(t)}
{\int_{0}^{1}dx\frac{x^2 (1-x)}{x^2
+{\frac{t}{m_{\mu^2}}}(1-x)}}\,,
\ee
with
\be
\sigma(t)_{e^+ e^-\ra {\mbox{\rm\tiny
hadrons}}}=\frac{4\pi^2\alpha}{t}{
\frac{1}{\pi}\Imm\Pi(t)}\,. 
\ee
The integration kernel in Eq.\rf{specrep} shows well the underlying
physical features. 
\begin{itemize}
\item The spectral function $\Imm\Pi(t)$, which is positive, is modulated
by a known function of $t$ which is also positive and monotonously
decreasing. The integral is therefore positive and dominated by the
low--energy region; mostly by the $\rho$ resonance.

\item In QCD, the spectral function at large--$t$ goes to a constant,
which ensures the UV convergence of the integral. Perturbative QCD fails,
however, to reproduce the observed hadronic shape of the spectral
function below $t\sim 1.5~\GeV^2$. In fact, perturbative QCD with
massless u and d quarks gives an IR--divergent result, pointing out the
importance of non--perturbative effects.

\item
There is a lower bound to the integral in Eq.\rf{specrep}
\be
a_{\mu}^{\mbox{\rm\tiny
(h.~v.p.)}}\ge \frac{\alpha}{\pi}\frac{1}{3}m_{\mu}^2 \int_{0}^{\infty}
dt\frac{1}{t^2}\Imm\Pi(t)\,,
\ee
which is governed by the slope of the hadronic vacuum polarization at
the origin\cite{BdeR69}.    
\end{itemize}

I have compiled in Table~1 the most recent
evaluations of the integral in Eq.\rf{specrep} at the time of the La
Thuile meeting, and I refer to the original literature for the details of
the various evaluations. While I was writing this talk, there 
appeared a new detailed evaluation\cite{DEHZ02}, which uses the recent
$e^{+} e^{-}$ data from the CMD-2 detector at Novosibirsk, as well as the
final analysis of hadronic $\tau$--decay from the ALEPH and CLEO
detectors at LEP. Unfortunately, the results found for
$a_{\mu}^{\mbox{\rm\tiny (h.~v.p.)}}$ from the $e^{+} e^{-}$--based data
and from the
$\tau$--based data are inconsistent with each other, even after
applying radiative corrections and isospin corrections:
\be\lbl{latest}
a_{\mu}^{\mbox{\rm\tiny
(h.~v.p.)}}=\left\{\begin{array}{lr}
(684.7\pm 6.0_{\mbox{\rm\tiny exp }}\pm 3.6_{\mbox{\rm\tiny rad }})\times
10^{-10} & [e^+ e^- -{\mbox{\rm\small based}}]\,,\\
(701.9\pm 4.7_{\mbox{\rm\tiny exp }} \pm 1.2_{\mbox{\rm\tiny rad }} \pm
3.8_{\mbox{\rm\tiny $SU(2)$ }})\times 10^{-10} & [\tau-{\mbox{\rm\small
based}}]\,.
\end{array}\right.
\ee    
\begin{table}[h]
\caption[]{\it Compilation of recent estimates from Hadronic Vacuum
Polarization}
\lbl{table1}
\begin{center}
\begin{tabular}{|c|c|} \hline \hline {\bf {Authors}} &
{\bf {Contribution to $a_{\mu}\times 10^{10}$}} 
\\ \hline \hline
{ Davier--H\"{o}cker\cite{DH}} & {$692.4 \pm 6.2$} 
\\
{ Jegerlehner\cite{Je}} & {$697.40 \pm 10.45$} \\ 
{ Narison \cite{Nar}} & {$702.06\pm 7.56$} \\
{ de Troc\'oniz--Yndur\'ain \cite{deTY}} &
{$695.2\pm 6.4$}
\\  
\hline\hline
\end{tabular}
\end{center}
\end{table}

Higher order hadronic vacuum polarization contributions were first
estimated in ref.\cite{CNPdeR77}. The most recent evaluation
in\cite{Kr97} gives the result
\be\lbl{hovp}
{a_{\mu}^{\mbox{\rm\tiny
(h.o.-h.~v.p.)}}=-10.0~(0.6)\times 10^{-10}}\,.
\ee
Concerning this evaluation, one should realize that there may be here a
potential problem  of double counting with some of the
lowest order estimates . This
is because {\it part} of the radiative hadronic corrections of the type
indicated by the diagram in Fig.~7 below have already been included in
the experimental cross section used to evaluate the {\it lowest order}
contribution in Fig.~6.
%%%%%%%%%%%%%%%%%%%%%%%%%%%%%%%%%%%%%%
\vskip 1pc
\centerline{\epsfbox{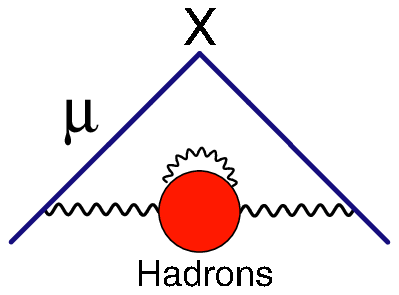}}
\begin{center}
{{\bf Fig.~7} {\it Higher Order Hadronic Vacuum Polarization
Contribution}}
\end{center}
%\vskip 1pc
%%%%%%%%%%%%%%%%%%%%%%%%%%%%%%%%%%%%%%
This issue of double counting is under
investigation at present\cite{KNPdeR}.

%%%%%%%%%%%%%%%%%%%%%%%%%%%%%%%%%%%%%%%%%%%%%%
\section{Hadronic Light--by--Light Scattering}
%%%%%%%%%%%%%%%%%%%%%%%%%%%%%%%%%%%%%%%%%%%%%%

\noi
These are the contributions illustrated by the diagrams in Fig.~8 below
%%%%%%%%%%%%%%%%%%%%%%%%%%%%%%%%%%%%%%
\vskip 1pc
\centerline{\epsfbox{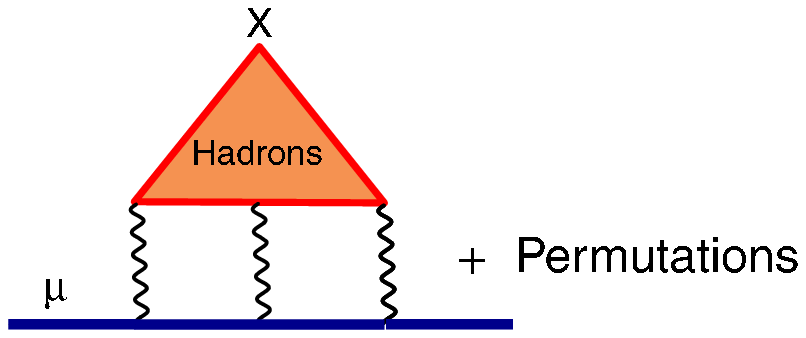}}
\begin{center}
{{\bf Fig.~8} {\it Hadronic Light--by--Light Contributions}}
\end{center}
%\vskip 1pc
%%%%%%%%%%%%%%%%%%%%%%%%%%%%%%%%%%%%%%
\noi
All the estimates of these contributions made so far are model dependent.
There has been progress, however, in identifying the dominant regions of
virtual momenta, and in  using models which incorporate some of the
required features of the underlying QCD theory. The combined frameworks
of QCD in the $1/N_c$--expansion and of chiral perturbation
theory\cite{deR94} have been very useful in providing a
guiding line to possible estimates.

Recent progress in this domain has come from the
observation\cite{KNPdeR02} that, in large--$N_c$ QCD and to leading order
in the chiral expansion, the dominant contribution to the muon $g-2$ from
hadronic light--by--light scattering comes from the contribution of the
diagrams which are one particle (Goldstone--like) reducible; these are
the diagrams in Fig.~9 below.
The first of these diagrams (Fig.~9a) produces a
$\log^2\left(\mu/m\right)$--term with a coefficient which is an {\it an
exact QCD result}:
\be\lbl{piefft}
a_{\mu}^{(\pi^{0})}=\left(\frac{\alpha}{\pi}
\right)^3\left\{{\frac{N_c^2}{48\pi^2}
\frac{m_{\mu}^2}{F_{\pi}^2}}\log^2\left(\frac{{\mu}}{m} \right) +
\cO\left[\log\left(\frac{{\mu}}{m}\right)+\kappa(\mu)
\right]\right\}\,.
\ee
Here, $F_{\pi}$ denotes the pion coupling constant in the chiral limit
($F_{\pi}\sim 90~\MeV$);  the
$\mu$--scale in the logarithm is an arbitrary UV--scale, and
$m$ an infrared mass (either $m_{\mu}$ or $m_{\pi}$).
%%%%%%%%%%%%%%%%%%%%%%%%%%%%%%%%%%%%%%
\vskip 1pc
\centerline{\epsfbox{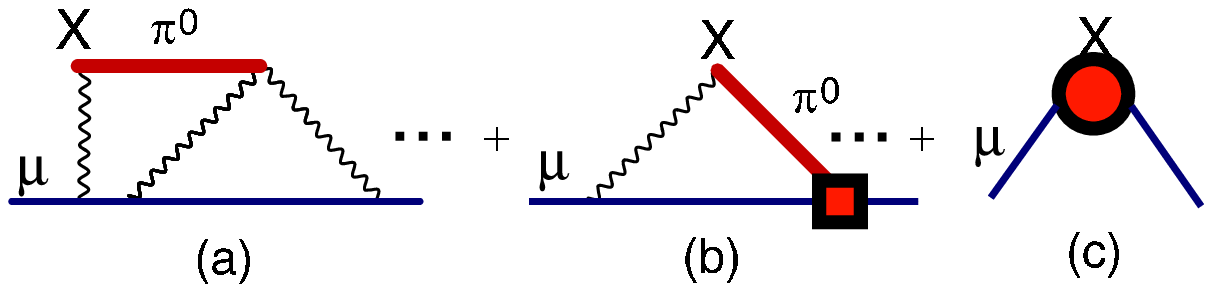}}
\begin{center}
{{\bf Fig.~9} {\it One Goldstone Reducible Diagrams in Chiral
Perturbation Theory}}
\end{center}
%\vskip 1pc
%%%%%%%%%%%%%%%%%%%%%%%%%%%%%%%%%%%%%%
The
dependence on the $\mu$--scale in Eq.\rf{piefft} would be removed, if one
knew the terms linear in $\log\mu$ from Fig.~9b, as well as the constant
$\kappa(\mu)$ from the local counterterms generated by Fig.~9c.
Unfortunately, the determination of some of the coefficients of the
$\log\mu$--terms cannot be made in a completely model independent
way\footnote{Ref.\cite{KNPdeR02} provides a discussion of this point
using a renormalization group approach. Essentially the same arguments
have been recently emphasized in ref.\cite{RW02}.}; neither the
determination of the constant $\kappa(\mu)$. Nevertheless,
Eq.\rf{piefft} plays a fundamental role in fixing the overall sign
of the hadronic light--by--light scattering contribution to the muon
$g-2$. In the various hadronic model calculations of this contribution,
there appear indeed hadronic scales (usually the $\rho$--mass), which act
as an UV--regulator, and  play the role of $\mu$ in Eq.\rf{piefft}.
Therefore, letting the hadronic scale become large, and provided that
the model incorporates correctly the basic chiral properties of
the underlying QCD theory, must reproduce the characteristic
universal
$\log^2(\mu)$ behaviour of Eq.\rf{piefft}, with the \underline{same}
coefficient. This test, when applied to the most recent existing
calculations\cite{HK98,BPP96} (prior to the Knecht--Nyffeler calculation
in ref.\cite{KNb01}) {\it failed} to reproduce the sign of the coefficient
of the
$\log^2(\mu)$--term in Eq.\rf{piefft}, though the results from the
calculations, when extrapolated to large UV--scales, agreed in absolute
value with the coefficient of the
$\log^2(\mu)$--term. The authors of refs.\cite{HK98,BPP96} have later
found mistakes in their calculations which, when corrected, reproduce the
effective field theory test. Their results, now, agree with the
Knecht--Nyffeler calculation\cite{KNb01} which we report on next.

\begin{description}

\item[]{\it The Knecht--Nyffeler Calculation}

In full generality, the pion pole contribution to the muon anomaly has
hadronic structure, as represented by the shaded blobs in Fig.~10 below.
The authors of ref.\cite{KNb01} have shown that, for a large class of
off--shell $\pi^{0}\gamma\gamma$ form factors (which includes the
large--$N_c$ QCD class), the contribution from these diagrams has an
integral representation over two euclidean invariants $Q_1^2$ and
$Q_2^2$ associated with the two loops in Fig.~10:
\be
a_{\mu}^{(\pi^{0}\mbox{\rm\tiny
~l.~by~l.})}=\int_{0}^\infty \!\!dQ_{1}^2\int_{0}^{\infty}\!\!
dQ_{2}^2\ \ \cW(Q_{1}^2,Q_{2}^2)
\cH(Q_{1}^2,Q_{2}^2)\,,
\ee
where $\cW(Q_{1}^2,Q_{2}^2)$ is a skeleton kernel which they
calculate explicitly, and   
$\cH(Q_{1}^2,Q_{2}^2)$ is a convolution of
two generic 
$\cF_{\pi^{0}\gamma^{*}\gamma^{*}}(k_{1}^2,k_{2}^2)$
form factors. In Large--$N_c$ QCD,
\be\lbl{pigg}
\cF_{\pi^{0}\gamma^{*}\gamma^{*}}
(k_{1}^2,k_{2}^2)\big\vert_{N_c\ra\infty}=\sum_{ij}
\frac{c_{ij}(k_{1}^2,k_{2}^2)}{(k_{1}^2-M_{i}^2)(k_{2}^2-M_{j}^2)}\,,
\ee
with the sum extended to an infinite set of narrow states. 
%%%%%%%%%%%%%%%%%%%%%%%%%%%%%%%%%%%%%%
\vskip 1pc
\centerline{\epsfbox{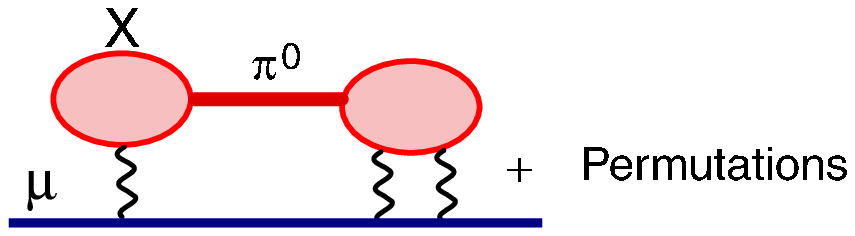}}
\begin{center}
{{\bf Fig.~10} {\it Hadronic Light--by--Light from a $\pi^{0}$ State}}
\end{center}
%\vskip 1pc
%%%%%%%%%%%%%%%%%%%%%%%%%%%%%%%%%%%%%%

In practice, the calculation in\cite{KNb01} has been done by restricting
the sum in Eq.\rf{pigg} to one and two vector states, and fixing the
polynomial
$c_{ij}(k_{1}^2,k_{2}^2)$ from general short--distances and
long--distances QCD properties. This way, they obtain the result
\be
{a_{\mu}^{(\pi^{0}\mbox{\rm\tiny
~l.~by~l.})}=(5.8 \pm 1.0)\times 10^{-10}}\,,
\ee
where the error also includes an estimate of the hadronic approximation.
Further inclusion of the $\eta$ and $\eta\prime$ states results in a
final estimate
\be
{a_{\mu}^{(\pi^{0}+\eta+\eta\prime\mbox{\rm\tiny
~l.~by~l.})}=(8.3 \pm 1.2)\times 10^{-10}}\,. 
\ee

\item[]{\it A Remark on the Constituent Quark Model (CQM)}

This is perhaps a good place to comment on an argument which is often used
in favor of the {\it constituent quark model} as a {\it simple} way to
{\it estimate} the hadronic light--by--light scattering contribution to
the muon $g-2$. Since the argument has even appeared in
print in some recent papers, I feel obliged to abrogate it here, so as to
stop further confusion.

The constituent quark model contribution from the diagram in Fig.~11
below,
%%%%%%%%%%%%%%%%%%%%%%%%%%%%%%%%%%%%%%
\vskip 1pc
\centerline{\epsfbox{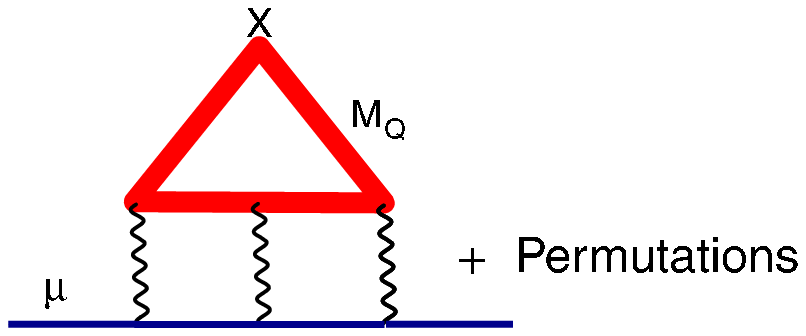}}
\begin{center}
{{\bf Fig.~11} {\it Hadronic Light--by--Light in the Constituent Quark
Model}}
\end{center}
%\vskip 1pc
%%%%%%%%%%%%%%%%%%%%%%%%%%%%%%%%%%%%%%
can be easily extracted from the work of Laporta and Remiddi in
ref.\cite{LR93}, with the result
\be
a_{\mu}^{{\mbox{\rm\tiny
(CQM)}}}\!=\!\left(\frac{\alpha}{\pi}
\right)^3\!N_{c}{\frac{2}{9}}
\left\{{\underbrace{\left[\frac{3}{2}\zeta(3)
-\frac{19}{16}
\right]}_{0.616} }\left(\frac{m_{\mu}}{{M_{Q}}}
\right)^2 \!\!+\!\!\cO\left[
\left(\frac{m_{\mu}}{{M_{Q}}}
\right)^4 \!\log^2 \left(\frac{{M_{Q}}}{m_{\mu}}
\right) \right]\right\}\,.
\ee
Seen from a low energy effective field theory point of view, the
constituent quark mass
$M_Q$ in the CQM should provide the UV--regulating scale. However, the
model {\it is not} a good effective theory of QCD and, therefore, it fails
to reproduce the characteristic QCD
$\log^2 M_Q$ behaviour when $M_Q$ is allowed to become arbitrarily large;
in fact the CQM result above, decouples in the large $M_Q$--limit. The
argument of a {\it positive} contribution based on the CQM is certainly a
{\it simple} argument, but unfortunately it is {\it wrong}. 
\end{description}

Notice however that, contrary to the naive CQM, the constituent
chiral quark model of Georgi and Manohar\cite{GM84} (see also
ref.\cite{EdeRT90}) does indeed reproduce the correct $\log^2 M_Q$
behaviour in the
$M_Q\ra\infty$ limit. This is because, in this model, the  Goldstone
particles couple with the constituent quarks in a way which respects
chiral symmetry, and the pion pole diagram appears then explicitly. (The
same happens in the Nambu--Jona-Lasinio model as well as in its extended
version, the ENJL--model\cite{BBdeR93}). These models, however, suffer
from other diseases\footnote{See e.g. the discussion in
ref.\cite{PPdeR98}}, and therefore are not fully reliable to compute
the hadronic light--by--light scattering contribution. Hopefully, they
will be progressively amended so as to incorporate further and further QCD
features, in particular the short--distance constraints, following the
line discussed in refs.\cite{deR02,Pe02}, as already applied to the
evaluation of the one particle (Goldstone--like)  reducible diagrams in
ref.\cite{KNb01} reported above. This is why, at the moment, one can only
claim to know the hadronic light--by--light scattering contribution with
a {\it cautious} generous error, which takes into account these
uncertainties. While awaiting for further improvement, the value quoted
(at present)  by our group in Marseille, based on the combined work of
refs.\cite{HK98,BPP96} (appropriately corrected) and ref.\cite{KNb01}, is
\be
a_{\mbox{\rm\tiny hadronic}}^{(\mbox{\rm\tiny
light by light)}}=(8\pm
4)
\times 10^{-10}\,.
\ee

%%%%%%%%%%%%%%%%%%%%%%%%%%%%%%%%%%%%%%%%%%%%%%
\section{Electroweak Contributions}
%%%%%%%%%%%%%%%%%%%%%%%%%%%%%%%%%%%%%%%%%%%%%%

\noi
The contribution to the anomalous magnetic moment of the muon from the
electroweak Lagrangian of the Standard Model, at the one loop level,
originates in the three diagrams of Fig.~12 below,

%%%%%%%%%%%%%%%%%%%%%%%%%%%%%%%%%%%%%%
\vskip 1pc
\centerline{\epsfbox{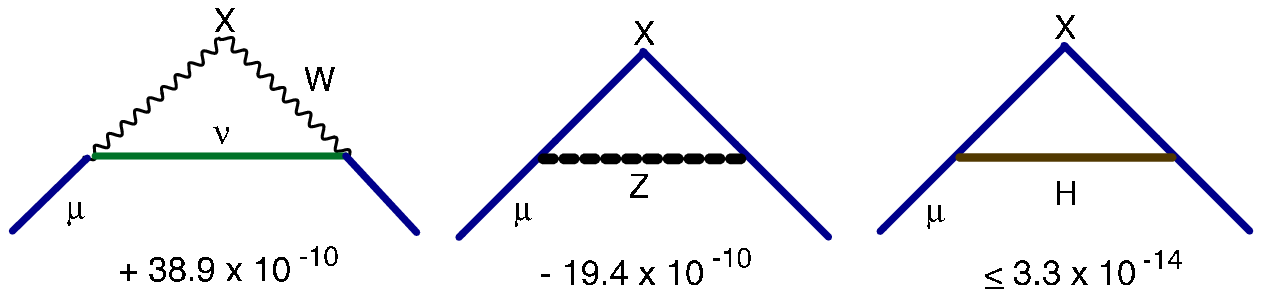}}
\begin{center}
{{\bf Fig.~12} {\it Weak Interactions at the one loop level}}
\end{center}
%\vskip 1pc
%%%%%%%%%%%%%%%%%%%%%%%%%%%%%%%%%%%%%%

\noi
where we also indicate
the size of their respective contributions. Their analytic evaluation
gives the result\cite{EW's}
 
{\setl
\bea
a_{\mu}^{\mbox{\rm\tiny
(W)}} & = &
\frac{G_{\mbox\tiny F}}{\sqrt{2}}\frac{m_{\mu}^2}{8\pi^2}
\left[{\frac{5}{3}\!+\!
\frac{1}{3}(1\!-\!4\sin^{2}\theta_{W})\!+\!\cO\left({
\frac{m_{\mu}^2}{M_{Z}^2}\log\frac{M_{Z}^2}{m_{\mu}^2}}\right)}\right.
\nn
\\
& & \left.
\hspace*{2.5cm}+{\frac{m_{\mu}^2}{M_{H}^2}}
\int_{0}^{1}dx 
\frac{2x^2
(2-x)}{1-x+{\frac{m_{\mu}^2}{M_{H}^2}}x^2}
\right]=19.48\times 10^{-10}\,. 
\eea}

\noi
Notice that the contribution from the Higgs decouples and is very
small. 

Let us recall that the present {\it world average} experimental error
in the determination of the muon anomaly is\cite{BNL1} 
$
\Delta a_{\mu}\big\vert_{\mbox{\rm\tiny Exp.}}\!\!\!=\pm 8\times
10^{-10}\,,
$
and, hoping for a continuation of the BNL experiment, it is expected to
be further reduced. A theoretical effort on
the evaluation of the two--loop electroweak corrections is therefore
justified. It is convenient to separate the two--loop electroweak contributions into
two sets of Feynman graphs: those which contain closed fermion loops,
which we denote by $a_{\mu}^{EW(2)}({\mbox{\rm\footnotesize ferm}})$,
and the others which we denote by 
$a_{\mu}^{EW(2)}({\mbox{\rm\footnotesize bos}})$. In this notation,
the electroweak contribution to the muon anomalous magnetic moment is
\be
a_{\mu}^{EW}=a_{\mu}^{W(1)}+a_{\mu}^{EW(2)}({\mbox{\rm\footnotesize
bos}})+a_{\mu}^{EW(2)}({\mbox{\rm\footnotesize ferm}})\,.
\ee
We shall review the calculation of the two--loop contributions
separately.

\begin{description}
\item[]{\it Bosonic Contributions}

The leading logarithmic terms of the two--loop electroweak bosonic
corrections have been extracted using asymptotic expansion
techniques. In fact, these contributions have
now been evaluated analytically, in a systematic expansion in powers of
$\sin^2\theta_{W}$, up to 
$\cO[(\sin^2\theta_{W})^3]\,,$ where $\log\frac{M_{W}^2}{m_{\mu}^2}$
terms, $\log\frac{M_{H}^2}{M_{W}^2}$ terms, $\frac{M_{W}^2}{M_{H}^2}
\log\frac{M_{H}^2}{M_{W}^2}$ terms, $\frac{M_{W}^2}{M_{H}^2}$ terms
and constant terms are kept~\cite{CKM96}. Using 
$\sin^2\theta_{W}=0.224$ and $M_{H}=250\,\GeV\,,$ the authors of
ref.\cite{CKM96} find
\be\lbl{bos}
a_{\mu}^{EW(2)}({\mbox{\rm\footnotesize
bos}})=\frac{\GF}{\sqrt{2}}\,\frac{m_{\mu}^2}{8\pi^2}\,
\frac{\alpha}{\pi}\times 
\left[-5.96\log\frac{M_{W}^2}{m_{\mu}^2}+0.19\right]=
\frac{\GF}{\sqrt{2}}\,\frac{m_{\mu}^2}{8\pi^2}\,
\left(\frac{\alpha}{\pi}\right)\times (-79.3)\,.
\ee

\item[]{\it Fermionic Contributions}

The discussion of the two--loop electroweak fermionic corrections is
more delicate. Because of the
cancellation between lepton loops and quark loops in the electroweak
$U(1)$ anomaly, in the diagrams in Fig.~13, one cannot separate
hadronic from leptonic effects any longer.
In fact, as discussed in
refs.\cite{PPdeR95,CKM95}, it is this cancellation which eliminates some
of the large logarithms which  were incorrectly kept in a previous
calculation in ref.\cite{KKSS92}. It is therefore appropriate to separate
the two--loop electroweak fermionic corrections into two classes: one is
the class arising from Feynman diagrams like in Fig.~13, with both
leptons and quarks in the VVA--triangle, including the graphs  where the
$Z$ lines are replaced by
$\Phi^{0}$ lines, if the calculation is done in the $\xi_{Z}$--gauge.  We
denote this class by
$a_{\mu}^{EW(2)}(l,q)\,.
$ The other class is defined by the rest of the diagrams, where quark
loops and lepton loops can be treated separately, which we call
$a_{\mu}^{EW(2)}({\mbox{\rm\footnotesize ferm-rest}})$
i.e.,
$$
a_{\mu}^{EW(2)}({\mbox{\rm \footnotesize
fer}})=a_{\mu}^{EW(2)}(l,q)+
a_{\mu}^{EW(2)}({\mbox{\rm\footnotesize ferm-rest}})\,.
$$
%%%%%%%%%%%%%%%%%%%%%%%%%%%%%%%%%%%%%%
\vskip 1pc
\centerline{\epsfbox{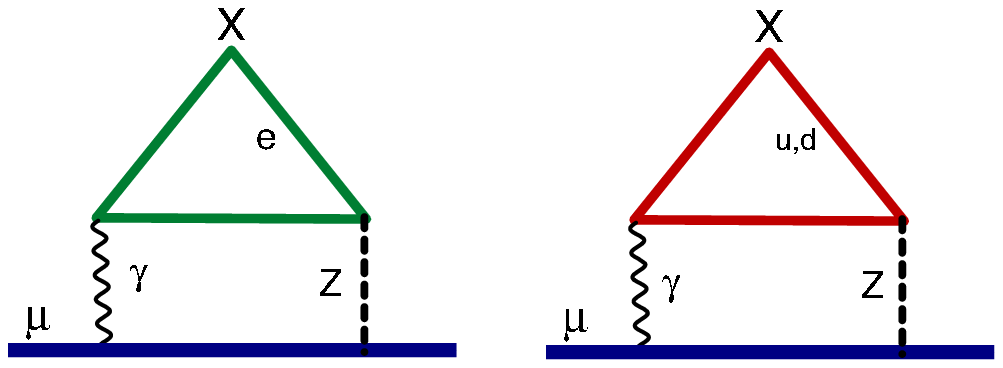}}
\begin{center}
{{\bf Fig.~13} {\it Two Loop Electroweak Diagrams generated by the
$\gamma\gamma Z$--Triangle}}
\end{center}
%\vskip 1pc
%%%%%%%%%%%%%%%%%%%%%%%%%%%%%%%%%%%%%%
The contribution from $a_{\mu}^{EW(2)}({\mbox{\rm\footnotesize
ferm-rest}})$ brings in $m_{t}^2/M_{W}^2$ factors. It has been
estimated, to a very good approximation, in ref.\cite{CKM95} with the
result,
\be\lbl{ferrest}
a_{\mu}^{EW(2)}({\mbox{\rm\footnotesize
ferm-rest}})=\frac{\GF}{\sqrt{2}}\,\frac{m_{\mu}^2}{8\pi^2}\,
\frac{\alpha}{\pi}\times\left(-21\,\pm\,4 \right)\,.
\ee

Concerning the contributions to
$a_{\mu}^{EW(2)}(l,q)$, it is convenient to treat
the contributions from the three generations separately. The
contribution from the third generation can be calculated in a
straightforward way, with the result~\cite{PPdeR95,CKM95}

{\setl
\bea\lbl{3rdg}
a_{\mu}^{EW(2)}(\tau,t,b) & = & \frac{\GF}{\sqrt{2}}\,\frac{m_{\mu}^2}
{8\pi^2}\,
\frac{\alpha}{\pi}  \times
 \left[-3\log\frac{M_{Z}^2}{m_{\tau}^2}-\log\frac{M_{Z}^2}{m_{b}^2}-
\frac{8}{3}\log\frac{m_{t}^2}{M_{Z}^2}+\frac{8}{3}\right. \nn \\
 &+ & \left. 
\cO\left(\frac{M_{Z}^2}{m_{t}^2}\log\frac{m_{t}^2}{M_{Z}^2}
\right)
\right] =\frac{\GF}{\sqrt{2}}\,\frac{m_{\mu}^2}
{8\pi^2}\,
\frac{\alpha}{\pi}\times (-30.6)\,.
\eea}

\noi
As emphasized in ref.\cite{PPdeR95}, an appropriate QCD calculation
when the quark in the loop of Fig.~1 is a {\it light quark} should
take into account the dominant effects of spontaneous chiral symmetry
breaking. Since this involves the $u$ and $d$ quarks, as well as the
second generation $s$ quark, it is convenient to lump together the
contributions  from the first and second generation. A
recent evaluation of these contributions\cite{KPPdeR02}, which
incorporates the QCD long--distance chiral realization as well as
short--distance constraints, gives the result
\be\lbl{12gs}
a_{\mu}^{EW(2)}(e,\mu,u,d,s,c)=\frac{\GF}{\sqrt{2}}\,\frac{m_{\mu}^2}
{8\pi^2}\,
\frac{\alpha}{\pi} \times (-28.5\pm 1.8)\,.
\ee
\end{description}

\noi
Putting together the numerical results in Eqs.\rf{bos}, \rf{ferrest},
\rf{3rdg} with the new result in Eq.\rf{12gs}, we finally obtain the
value
\be
a_{\mu}^{EW}\!=\!\frac{\GF}{\sqrt{2}}\,\frac{m_{\mu}^2}
{8\pi^2}\left[\frac{5}{3}\!+\!\frac{1}{3}
\left(1\!\!-\!\! 4\sin^2\theta_{W}\right)^2\!\!-\!\left(
\frac{\alpha}{\pi}\right) (159\!\pm\!4)\right]\!=\!(15.2\, \pm\,
0.1)\times 10^{-10}\,, 
\ee
which shows  that the two--loop correction represents a sizeable reduction
of the one--loop result by an amount of $22\%\,.$ 

%%%%%%%%%%%%%%%%%%%%%%%%%%%%%%%%%%%%%%%%%%%%%%%%%%%%%
\section{Summary of the Standard Model Contributions}
%%%%%%%%%%%%%%%%%%%%%%%%%%%%%%%%%%%%%%%%%%%%%%%%%%%%%

\noi
The situation, at present, concerning the evaluation of the anomalous
magnetic moment of the muon in the Standard Model, can be summarized as
follows:

\begin{itemize}
\item {\it Leptonic QED contributions}
$$a_{{\mbox{\rm\tiny
QED}}}(\mu)={11~658~470.57\pm 0.29\times 10^{-10}}
$$

\item {\it Hadronic Contributions}
\begin{itemize}
\item{\it Hadronic Vacuum Polarization}

It is clear that, given the present experimental accuracy in
Eq.(\ref{wa}), we need now a better understanding of the hadronic vacuum
polarization contributions. Issues like the possible double counting
already mentioned, and the improvement in the treatment of
radiative corrections and isospin corrections\cite{CEN02,HGJ02} have now
become extremely important.  For reference, I shall choose the
two results from the most recent determination in Eq.~\rf{latest},
combined with the higher order vacuum polarization estimate in
Eq.~\rf{hovp}.

\item {\it Hadronic Light--by--Light Scattering}
$$a_{\mbox{\rm\tiny hadronic}}^{(\mbox{\rm\tiny
light by light)}}=\underbrace{(8\pm
4)}_{{\mbox{\rm\tiny work in
progress}}}
\times 10^{-10}$$
\end{itemize}
\item {\it Electroweak Contributions }
$$a_{\mbox{\rm\tiny EW}}=(15.2\pm
0.1)\times 10^{-10} $$
\end{itemize}

\noi
The sum of these contributions, adding experimental and theoretical
errors in quadrature, gives then a total
\be
a_{\mu}^{\mbox{\rm\tiny SM}}=\left\{\begin{array}{lr} (11~659~168.5\pm
8.1)\times 10^{-10} & [e^+e^- -{\mbox{\rm\small based }}]\,, \\ 
(11~659~185.7\pm 7.4)\times
10^{-10}  & [\tau-{\mbox{\rm\small based }}]\,.\end{array}\right.
\ee
to be compared to the experimental world average in Eq.~(\ref{wa})
$$
a_{\mu}^{\mbox{\rm\tiny
exp}}~=(11~659~203\pm 8)\times
10^{-10}\,. 
$$
Therefore, with the input for the Standard Model contributions discussed
above, one finds:

\begin{center}
\framebox{$a_{\mu}^{\mbox{\rm\tiny
exp}}-a_{\mu}^{\mbox{\rm\tiny SM}}=\left\{ \begin{array}{lcr}
(34.5\pm 11.4)\times
10^{-10} &  {3.0\sigma\ {\mbox{\rm discrepancy}}} & 
[e^+e^- -{\mbox{\rm\small based }}]\,, \\
(17.3\pm 10.9)\times
10^{-10} &  {1.6\sigma\ {\mbox{\rm discrepancy}}} & 
[\tau-{\mbox{\rm\small based }}]\,. \end{array}\right.
$}
\end{center}

\noi
We should be prepared for a new avalanche of theoretical speculations!

\vspace*{1cm}
\noi
{\bf Acknowledgements}

\noi
I wish to thank my colleagues Marc Knecht, Santi Peris, Michel
Perrottet and Andreas Nyffeler for a very pleasant collaboration on the
topics reported here. This work has been partly supported by the TMR, 
EC--Contract No.
ERBFMRX-CT980169 (EURODA$\Phi$NE).

%%%%%%%%%%%%%%%%%%%%%%%%%%%
%%%%%%%%%%%%%%%%%%%%%%%%%%%
%%%%%%%%%%%%%%%%%%%%%%%%%%%

%
\end{document}